\documentclass[%
 reprint,
 amsmath,amssymb,
 aps,
 superscriptaddress,
]{revtex4-2}
\usepackage{hyperref}
\usepackage{graphicx}
\usepackage{dcolumn}
\usepackage{bm}
\usepackage{braket} 
\usepackage{xcolor} 
\usepackage{hyperref}
\hypersetup{
    colorlinks=true,
    allcolors=blue
}

\begin{document}

\preprint{....}

\title{Observable-Level Framework for Finite-Key Security in Coherent-One-Way Quantum Key Distribution}

\author{Aryan Kuntal}
\email{aryan.kuntal09@gmail.com}
\affiliation{Center for Quantum Science and Technologies, Indian Institute of Technology Mandi, Kamand 175075, Himachal Pradesh, India}
\affiliation{Fergusson College, Gopal Krishna Gokhale Road, Shivajinagar, Pune 411004, Maharashtra, India}

\author{Utkarsh Giradkar}
\email{utkarshgiradkar1729@gmail.com}
\affiliation{Center for Quantum Science and Technologies, Indian Institute of Technology Mandi, Kamand 175075, Himachal Pradesh, India}

\author{Mudit Chaudhary}
\affiliation{Center for Quantum Science and Technologies, Indian Institute of Technology Mandi, Kamand 175075, Himachal Pradesh, India}
\affiliation{School of Physical Sciences, Indian Institute of Technology Mandi, Kamand 175075, Himachal Pradesh, India}

\author{Amit D. Lad}
\email{amitdlad@iitmandi.ac.in}
\affiliation{Center for Quantum Science and Technologies, Indian Institute of Technology Mandi, Kamand 175075, Himachal Pradesh, India}

\date{\today}

\begin{abstract}

Practical security analysis of Coherent-One-Way (COW) quantum key distribution requires optical-device modeling, finite experimental statistics, and security inference to be treated consistently without conflating their respective assumptions. We present a modular observable-level simulation framework organized through the pipeline $\boldsymbol{\Theta}_{\rm phys}
\xrightarrow{\mathcal P}\mathbf p
\xrightarrow{\mathcal G}\mathcal O
\xrightarrow{\mathcal S_j}\mathcal S_j(\mathcal O)$, in which physical source, channel, receiver, and detector parameters determine detector-resolved event probabilities, a finite-block layer applies an effective dead-time treatment and generates aggregate detector counts, and interchangeable security evaluators act on the same observable record. The framework is verified against analytical optical, virtual-state, and finite-key relations and evaluated against recently reported experimental COW-QKD operating points. The Kato finite-key calculation reproduces the refined experimentally reported secure-key rates most closely at intermediate distances, with deviations of approximately $-1.5\%$ at 50~km and $-10.3\%$ at 75~km. Monitoring bright-port decoy counts agree within approximately $0.9$--$2.4\%$, while the remaining distance-dependent dark-port residual identifies the principal model-sensitive observable of the present physical description. Applying Kato and Azuma--Hoeffding concentration treatments to identical detector records isolates the effect of finite-key statistical inference and shows substantially tighter Kato phase-error bounds for the rare-event monitoring records studied here. Holding the Kato evaluator fixed, joint optimization of the coherent-state mean photon number $\mu$ and receiver splitting fraction $t_B$ increases the model-predicted secure-key rate by $7.4\%$--$20.2\%$ relative to optimization of $\mu$ alone with $t_B=0.3$. The framework therefore provides a common computational interface for experimental reproduction, controlled security-model comparison, model-based protocol optimization, and modular adversarial stress testing in COW-QKD.

\end{abstract}

\maketitle

\section{Introduction}
\label{sec:introduction}

Translating quantum key distribution (QKD)~\cite{RevModPhys.81.1301,lo_secure_2014,diamanti_practical_2016,Pirandola:20} into practical optical systems requires source statistics, channel loss, detector nonidealities, and security-proof assumptions to be treated consistently~\cite{gottesman_security_2004}. In the finite-key regime, experimentally observed statistics must additionally be converted into security statements with explicitly controlled failure probabilities~\cite{renner_koenig_2005,renner_thesis_2005,tomamichel_finite_2012,tomamichel_quantum_2016}. Simulation therefore provides a natural route for tracing how experimentally accessible detector observables propagate into finite-key security estimates.

The Coherent-One-Way (COW) protocol~\cite{stucki_fast_2005} is particularly well suited to this problem. Logical information is encoded in the time of arrival of weak coherent pulses, while inter-pulse phase coherence is monitored to constrain an eavesdropper's information~\cite{huttner_quantum_1995,brassard_limitations_2000}. The architecture supports high-speed~\cite{stucki_continuous_2009}, long-distance~\cite{korzh_provably_2015}, and integrated-photonic~\cite{sibson_chip_2017} implementations. However, COW security cannot be inferred from the data-line quantum bit error rate alone: monitoring-line decoy statistics are required to bound unobserved complementary-basis quantities~\cite{gao_simple_2022,li_finite-key_2024}. Consequently, finite-key performance is particularly sensitive to sparse monitoring events and their statistical fluctuations~\cite{li_finite-key_2024,cao_experimental_2026}.

In finite-key COW-QKD, the sparse monitoring records must be converted into phase-error estimates with explicitly controlled failure probabilities. Azuma--Hoeffding provides a general martingale concentration bound based on bounded increments, whereas Kato's inequality can additionally exploit information from the observed event count. For the low-count monitoring regime considered here, this count-adaptive treatment can yield substantially tighter phase-error estimates and higher certified key rates than the Azuma--Hoeffding treatment, without implying a universal ordering of the two concentration inequalities \cite{li_finite-key_2024,PhysRevResearch.5.023132,kato_concentration_2020}.

Security analyses of COW have developed along complementary directions. Upper-bound and adversarial studies have examined zero-error attacks, quantum soft filtering, unambiguous-state-discrimination strategies, and related security limitations~\cite{branciard_zero-error_2007,branciard_upper_2008,gonzalez-payo_upper_2020,trenyi_zero-error_2021,kronberg_soft_filtering_2020,reydominguez_hacking_cow_2024,lutkenhaus_quantum_2002,avanesov_generalized_pns_2025,curty_foiling_zero_error_2021,lavie_improved_2022}. Lower-bound analyses have progressed from virtual entanglement-based formulations~\cite{gao_simple_2022} to composable finite-key treatments employing concentration inequalities such as Kato's bound~\cite{li_finite-key_2024}, enabling recent long-distance experiments~\cite{cao_experimental_2026,pathania_finite_2026}. For the sparse monitoring records encountered in COW-QKD, Kato's count-adaptive treatment can provide substantially tighter finite-key estimates than the more conservative Azuma--Hoeffding approach.

Numerical simulation of QKD is itself well established. General-purpose frameworks~\cite{fan-yuan_universal_2020,NuQKD,caputo_coherent_simulator_2022} and COW-specific simulators~\cite{khaleel_cow_simulation_2012,roy_cow_2024,brazaola_cow_simulator_2026} primarily emphasize optical, system-level, or engineering performance, whereas finite-key security proofs are formulated in terms of particular detector gains and finite counts. This motivates an explicit intermediate representation that separates the experimentally accessible detector record from both the internal physical model and the subsequently selected security treatment.

\begin{figure*}[t]
    \centering
    \includegraphics[width=0.96\textwidth]{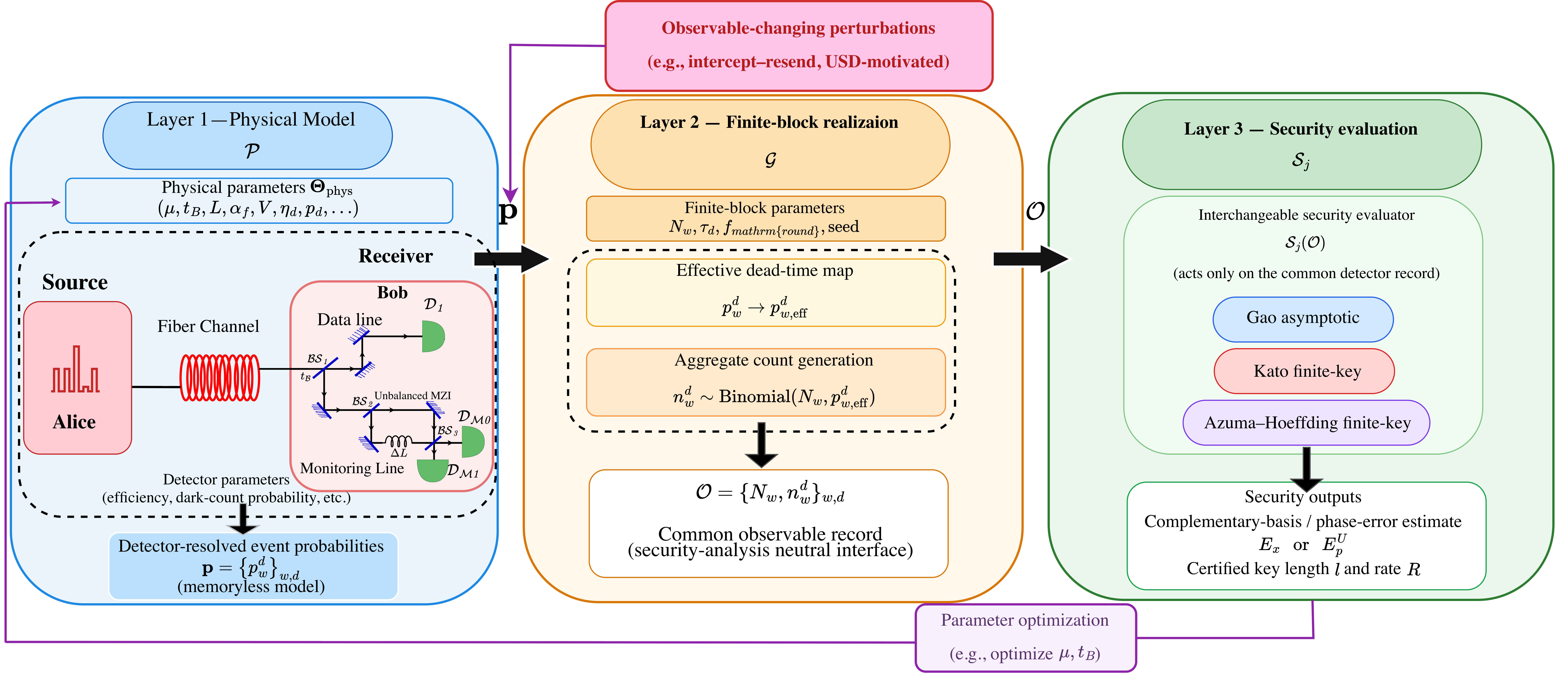}
    \caption{Observable-level architecture of the finite-key COW-QKD framework. Layer~1 maps the physical source, channel, receiver, and memoryless detector parameters to the detector-resolved event-probability vector $\mathbf p$. Within Bob's receiver, BS1 directs a fraction $t_B$ of the received field to the data detector $D_T$, while the complementary output enters the monitoring line. The monitoring measurement is represented in the model by an equivalent unbalanced two-path interferometer formed by BS2 and BS3, producing the outputs $D_{M0}$ and $D_{M1}$. The experimental receiver of Cao \textit{et al.} implements the corresponding monitoring measurement using an asymmetric Michelson interferometer with Faraday mirrors \cite{cao_experimental_2026}. Layer~2 applies the effective non-paralyzable dead-time map and aggregate finite-block count generation to construct the common observable record $\mathcal O$. Layer~3 acts only on $\mathcal O$, allowing the Gao asymptotic, Kato finite-key, and Azuma--Hoeffding finite-key evaluators to infer the complementary-basis quantity or finite-key phase-error upper bound and the corresponding certified secret-key length and rate. The observable-changing perturbation pathway acts upstream of the security evaluator, whereas the optimization feedback modifies the physical operating parameters. This separation enables experimental benchmarking, same-record security comparison, security-aware parameter optimization, and modular adversarial stress testing.}
    \label{fig:Schematic}
\end{figure*}

We therefore introduce the observable-level simulation architecture illustrated in Fig. ~\ref{fig:Schematic},
\begin{equation}
\boldsymbol{\Theta}_{\rm phys}
\xrightarrow{\mathcal P}
\mathbf p
\xrightarrow{\mathcal G}
\mathcal O
\xrightarrow{\mathcal S_j}
\mathcal S_j(\mathcal O).
\label{eq:pipeline}
\end{equation}
where $\mathcal P$ maps the physical source, channel, receiver, and detector parameters $\boldsymbol{\Theta}_{\rm phys}$ to the memoryless detector-resolved event-probability vector $\mathbf p$, $\mathcal G$ generates the finite experimental record $\mathcal O$, and $\mathcal S_j$ denotes an interchangeable security evaluator acting only on $\mathcal O$. This separation allows the physical backend to be refined without altering the downstream security interface, enables different finite-key treatments to be compared on the identical detector record, and permits protocol optimization or observable-changing perturbations to propagate through the same computational pipeline.

We verify the individual mappings against analytical relations and benchmark the complete framework against the experimental operating points of Cao \textit{et al.}~\cite{cao_experimental_2026}. Using identical finite detector records, we isolate the difference between Kato and Azuma--Hoeffding finite-key inference and optimize the secure-key-rate landscape over the coherent-state mean photon number $\mu$ and receiver splitting fraction $t_B$. Representative intercept--resend~\cite{gisin_towards_2004} and USD-motivated~\cite{branciard_zero-error_2007} perturbations are additionally used as modular stress tests. The objective is not a new COW security proof, but a common computational interface linking physical observables, finite experimental statistics, and interchangeable security inference.

\section{OBSERVABLE-LEVEL FRAMEWORK FOR COW-QKD}
\label{sec:methodology}

\subsection{Security-Relevant COW Observables}
\label{sec:cow_protocol}
We adopt the modified COW state-preparation and receiver configuration introduced by Gao \textit{et al.}~\cite{gao_simple_2022}, treated in the finite-key regime by Li \textit{et al.} \cite{li_finite-key_2024}, and implemented experimentally by Cao \textit{et al.}~\cite{cao_experimental_2026}. Each preparation round occupies two consecutive temporal modes, $2k-1$ and $2k$. We consider the modified prepare-and-measure formulation of Gao \textit{et al.} \cite{gao_simple_2022}, treated in the finite-key regime by Li \textit{et al.} \cite{li_finite-key_2024} and implemented experimentally by Cao \textit{et al.} \cite{cao_experimental_2026}. 

Each preparation round occupies two consecutive temporal modes, $2k-1$ and $2k$. Alice prepares coherent states of amplitude $\alpha$ with mean photon number $\mu = |\alpha|^2$, randomly selecting one of four two-mode preparations: signal states
\begin{align}
    |0_z\rangle &= |0\rangle_{2k-1}|\alpha\rangle_{2k}, \\
    |1_z\rangle &= |\alpha\rangle_{2k-1}|0\rangle_{2k},
\end{align}
or decoy preparations $|\alpha\alpha\rangle = |\alpha\rangle_{2k-1}|\alpha\rangle_{2k}$ and $|00\rangle = |0\rangle_{2k-1}|0\rangle_{2k}$ with respective probabilities $P_{\rm sig}$, $P_{\alpha\alpha}$, and $P_{00}$ summing to unity \cite{gao_simple_2022}. 

Bob passively divides the received field via an asymmetric beamsplitter of transmittance $t_B$ into a data line and a monitoring line \cite{cao_experimental_2026}. 
\begin{itemize}
    \item \textbf{Data-line observables:} Time-of-arrival detection uses threshold detector $D_T$ ($T_0 \rightarrow 1$, $T_1 \rightarrow 0$) \cite{cao_experimental_2026}. The empirical gain is $\widehat Q_w^d = n_w^d / N_w$, and the directly observed $Z$-basis bit-error rate is $E_z = (n_{0_z}^{T_0} + n_{1_z}^{T_1})/n_z$. 
    \item \textbf{Monitoring-line observables:} An unbalanced two-path interferometer with a one-bin delay measures the inter-pulse coherence using detectors $D_{M0}$ (bright port) and $D_{M1}$ (dark port). For the physical model, this measurement is represented by an equivalent Mach--Zehnder transformation; the experimental implementation of Cao \textit{et al.} employs an asymmetric Michelson interferometer with Faraday mirrors \cite{cao_experimental_2026}. The monitoring visibility is $V_{\alpha\alpha} = (\widehat Q_{\alpha\alpha}^{M_0} - \widehat Q_{\alpha\alpha}^{M_1})/(\widehat Q_{\alpha\alpha}^{M_0} + \widehat Q_{\alpha\alpha}^{M_1})$ \cite{gao_simple_2022}. 
    \item \textbf{Observable record:} The security-analysis-neutral record is defined as $\mathcal O=\{N_w,n_w^d\}_{w,d}$ for $w\in\{0_z,1_z,\alpha\alpha,00\}$. 
\end{itemize}

\subsection{Virtual-State Formulation and Phase-Error Inference}
\label{sec:eb_description}

Following Gao \textit{et al.}~\cite{gao_simple_2022}, security is evaluated through an equivalent entanglement-based description. Alice prepares
\[ |\Psi\rangle_{AA'} = \frac{1}{\sqrt{2}} \left(|0\rangle_A|0_z\rangle_{A'} + |1\rangle_A|1_z\rangle_{A'}
\right), \] and a virtual $X$-basis measurement projects the optical system onto the orthogonal states $|0_x\rangle$ and $|1_x\rangle$ with probabilities $N^{\pm}/4$, where \[ N^{\pm}=2(1\pm e^{-\mu}). \]

The corresponding virtual $X$-basis error rate is related to the monitoring statistics by~\cite{gao_simple_2022}
\begin{equation}
E_x =
\frac{
N^+ Q_{0_x}^{M_1}
+
2\left(Q_{0_z}^{M_0}+Q_{1_z}^{M_0}\right)
-
N^+ Q_{0_x}^{M_0}
}{
2\left(
Q_{0_z}^{M_0}
+
Q_{0_z}^{M_1}
+
Q_{1_z}^{M_0}
+
Q_{1_z}^{M_1}
\right)
}.
\end{equation}

The virtual gains $Q_{0_x}^{M_0}$ and $Q_{0_x}^{M_1}$ are not directly measurable. Decomposition of $|0_x\rangle$ introduces the physically prepared decoy components $|00\rangle$ and $|\alpha\alpha\rangle$, together with an auxiliary non-vacuum component $|\beta\beta\rangle$ that is not directly prepared~\cite{gao_simple_2022}. Positivity of the measurement operator and the Cauchy--Schwarz inequality then bound the virtual gains from the experimentally accessible decoy statistics, yielding $\underline{Q}_{0_x}^{M_0}$ and $\overline{Q}_{0_x}^{M_1}$.

Finite-key statistical estimation converts these quantities into a conservative phase-error upper bound $E_p^U$. The certified secret-key length is~\cite{cao_experimental_2026}
\begin{equation}
\ell = n_z\left[1-h_2\left(E_p^U\right)\right] - \mathrm{leak}_{\rm EC} - \log_2\frac{2}{\epsilon_{\rm cor}}
- 2\log_2\frac{5}{\epsilon_{\rm sec}},
\label{eq: secret_key}
\end{equation} where $\mathrm{leak}_{\rm EC}=f n_z h_2(E_z)$,
$f=1.1$, $\epsilon_{\rm cor}=10^{-15}$, and $\epsilon_{\rm sec}=10^{-10}$~\cite{cao_experimental_2026}; $h_2$ denotes the binary entropy function. 

\subsection{Modular Physical-to-Security Architecture}
\label{sec:simulation_architecture}
The simulation framework implements the three mappings introduced in Eq.~ \ref{eq:pipeline} as a layered pipeline that decouples continuous-variable optical propagation, stochastic discrete detection, and statistical security evaluation (Fig.~\ref{fig:Schematic}).

\subsubsection{Physical-to-Probability  Mapping ($\mathcal P$)}

Layer 1 maps the physical source, channel, receiver, and memoryless detector parameters $\boldsymbol{\Theta}_{\rm phys} =\{\mu,\alpha_f,L,t_B,V,\eta_d,p_d,\ldots\}$ to the detector-resolved event-probability vector $\mathbf p=\{p_w^d\}_{w,d}$. Coherent-state propagation is modeled using the continuous-variable Gaussian engine of Strawberry Fields~\cite{killoran_strawberry_2019,weedbrook_gaussian_2012}, avoiding Fock-space truncation and providing an exact representation within the adopted Gaussian optical model. Fiber attenuation over distance $L$ is represented by
$\eta_{\rm ch}=10^{-\alpha_f L/10}$.

At Bob, an asymmetric beam splitter of transmittance $t_B$ directs the received field to the data and monitoring lines. For data detector $D_T$, the mean detected photon number
$\lambda_{T,j}=\eta_T t_B\eta_{\rm ch}|\alpha_j|^2$
gives the threshold click probability
\begin{equation}
p_{T,j}=1-(1-p_{{\rm d},T})\exp(-\lambda_{T,j}).
\end{equation}
In the monitoring line, consecutive temporal modes interfere in an unbalanced two-path interferometer. In the present Gaussian model this measurement is represented by an equivalent Mach--Zehnder transformation, while the receiver of Cao \textit{et al.} realizes the corresponding measurement using an asymmetric Michelson interferometer \cite{cao_experimental_2026}. Residual phase fluctuations are modeled as $\delta\sim\mathcal N(0,\sigma_\phi^2)$, with $V=\exp(-\sigma_\phi^2/2)$. The resulting bright- and dark-port probabilities are averaged over this distribution using a 12-node Gauss--Hermite quadrature~\cite{abramowitz_handbook_1964}, providing a high-accuracy deterministic evaluation without additional Monte Carlo sampling.

\subsubsection{Finite-Block Observable Generation ($\mathcal G$)}
\label{sec:finite_block_model}

Layer 2 maps the detector-resolved event-probability vector $\mathbf p$ to the finite experimental record $\mathcal O$. Over an acquisition block, Alice emits \[ N_{\rm tot} = N_{0_z}+N_{1_z}+N_{\alpha\alpha}+N_{00} \] two-bin COW preparation rounds.

Detector recovery is incorporated through a stationary non-paralyzable dead-time model~\cite{michotte_deadtime_2009}. Each COW preparation round contains two consecutive temporal modes; therefore, the $500$~MHz optical pulse clock corresponds to a preparation-round rate
$f_{\rm round}=2.5\times10^{8}$~Hz. Since the Layer~1 probabilities $p_w^d$ are defined per two-bin preparation round, the effective click probability is
\begin{equation}
p_{w,\mathrm{eff}}^d
=
\frac{p_w^d}
{1+p_w^d f_{\rm round}\tau_d},
\end{equation}
where $\tau_d$ denotes the effective detector recovery time. Unless otherwise stated, we use $\tau_d=100$~ns as a model parameter in the stationary non-paralyzable dead-time approximation~\cite{michotte_deadtime_2009}.

For detector channels treated under this effective independent-event model, aggregate counts are generated as
\begin{equation}
n_w^d
\sim
\mathrm{Binomial}
\left(
N_w,p_{w,\mathrm{eff}}^d
\right),
\end{equation}
yielding the common observable record
\[ \mathcal O=\{N_w,n_w^d\}_{w,d}. \]
Sampling variability is summarized using Wilson score intervals~\cite{wilson_probable_1927}, while parameter searches use a fixed pseudorandom seed to avoid artificial objective-function roughness from independently resampled finite-block realizations. Because $N_{\rm tot}$ counts two-bin preparation rounds, per-round key quantities are converted to bits per second using $f_{\rm round}=250$~MHz.

\subsubsection{Interchangeable Security Evaluation ($\mathcal S_j$)}

Layer 3 operates exclusively on the observable record $\mathcal O$, leaving the physical model and finite-count generator unchanged. From $\mathcal O$, the evaluator computes the empirical gains $\widehat Q_w^d=n_w^d/N_w$ and the data-line bit-error rate $E_z$. The unmeasurable virtual-state gains $\underline{Q}_{0_x}^{M0}$ and $\overline{Q}_{0_x}^{M1}$ are bounded from the physical decoy statistics using the Cauchy--Schwarz construction of Gao \textit{et al.}~\cite{gao_simple_2022}.

Finite-size fluctuations are then bounded using either the Azuma--Hoeffding~\cite{azuma_weighted_1967} or Kato~\cite{kato_concentration_2020} concentration inequality. Azuma--Hoeffding provides a general martingale bound based on bounded increments, whereas Kato additionally exploits information from the observed event count to obtain count-adaptive deviations. This distinction can substantially tighten the bound for the sparse monitoring records encountered in COW-QKD~\cite{li_finite-key_2024}. Both evaluators act on the identical record $\mathcal O$, so differences in the resulting phase-error bound $E_p^U$ arise solely from the statistical inference. The certified secret-key length \[ \ell=\mathcal S_j(\mathcal O) \] is then obtained from Eq.~\eqref{eq: secret_key}.

\section{Experimental Grounding and Controlled Finite-Key Evaluation}
\label{sec:results}

\subsection{Verification of the Observable and Security Interfaces}
\label{sec:results_verification}

The internal consistency of the computational pipeline was verified sequentially across its three layers. The physical observable map $(\boldsymbol{\Theta}_{\rm phys} \xrightarrow{\mathcal P}\mathbf p)$ was verified against closed-form coherent-state threshold probabilities and the ideal-visibility COW monitoring relations~\cite{gao_simple_2022,weedbrook_gaussian_2012}. The finite-block generator $(\mathbf p\xrightarrow{\mathcal G}\mathcal O)$ was verified against the expected means and variances of the aggregate binomial count model. The security interface $(\mathcal O\xrightarrow{\mathcal S_j} \mathcal S_j(\mathcal O))$ reproduced the virtual-state Cauchy--Schwarz construction of Gao \textit{et al.}~\cite{gao_simple_2022} and the Kato finite-key bias parameters~\cite{kato_concentration_2020}. Additional implementation checks are provided in the Supplemental Material.

\begin{figure}[htbp]
    \centering
    \includegraphics[width=0.48\textwidth]{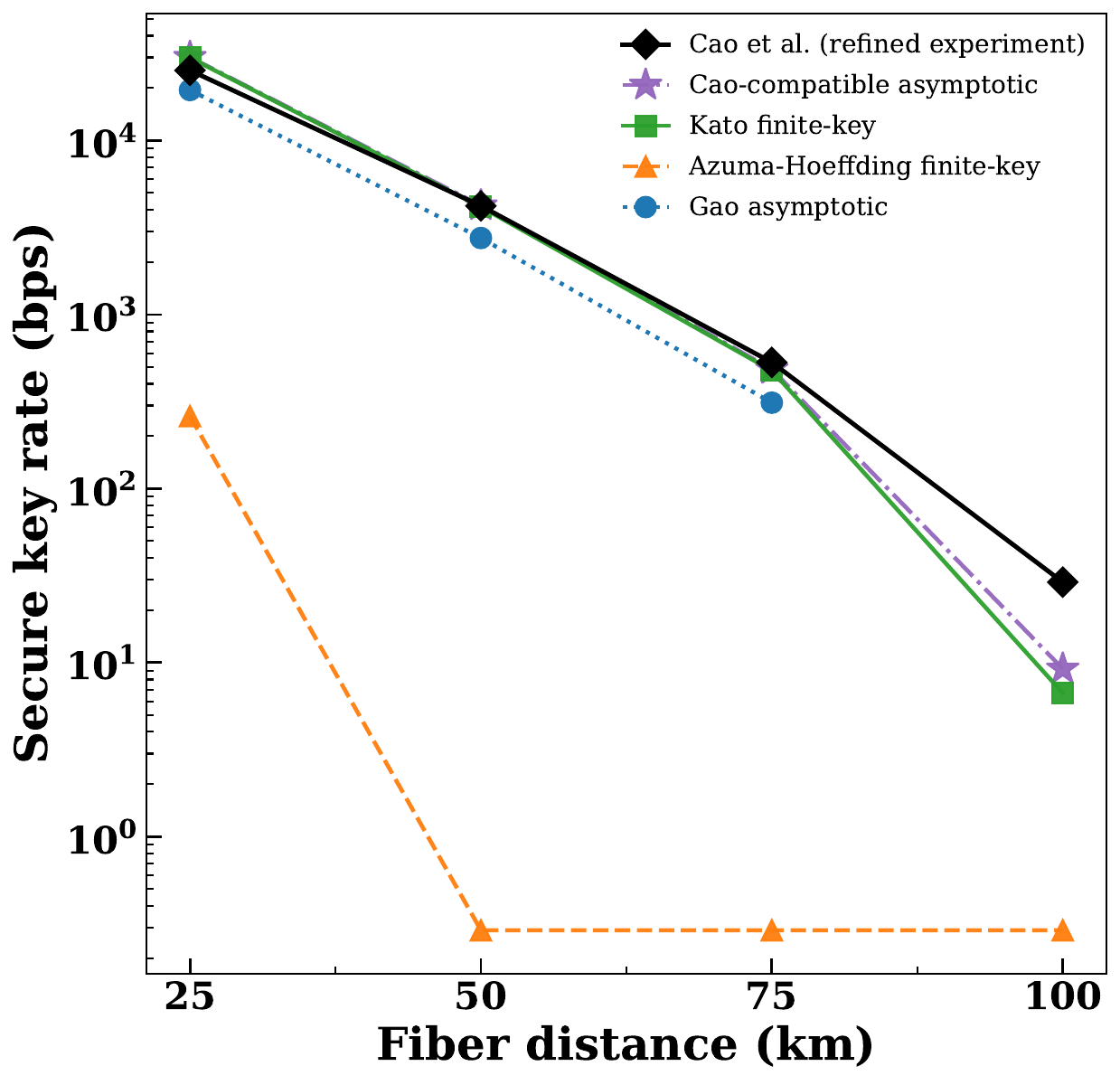}
    \caption{Secure-key-rate comparison against experimental benchmarks. The discrete markers denote the refined experimental secure-key rates reported by Cao \textit{et al.}~\cite{cao_experimental_2026}. The Cao-compatible asymptotic, Kato finite-key, and Azuma--Hoeffding finite-key results are evaluated using the corresponding reported operating parameters at each distance. The Gao asymptotic result is evaluated separately as an optimized reference under its own state-preparation convention and therefore does not constitute a like-for-like reproduction of the experimental operating points. Azuma--Hoeffding results are shown only at operating points for which the adopted finite-key bound certifies a positive secret key.}
\label{fig:experimental_reproduction}
\end{figure}

\subsection{Experimental Reproduction and Model Calibration}
\label{sec:experimental_benchmark}

Having verified the internal interfaces, the complete framework was evaluated at the experimental operating points reported by Cao \textit{et al.}~\cite{cao_experimental_2026}. Using the reported experimental parameters and an effective visibility $V_{\rm cal}=0.9964$ calibrated from the monitoring dark-port decoy statistics, the relative secure-key-rate residuals are $+17.3\%$, $-1.5\%$, $-10.3\%$, and $-76.9\%$ at $L=\{25,50,75,100\}$~km, respectively. Agreement is strongest at the intermediate distances, where the calculated rates differ from experiment by approximately $-1.5\%$ at 50~km and $-10.3\%$ at 75~km.

Importantly, the simulated monitoring bright-port decoy counts, which are not used to determine $V_{\rm cal}$, agree with the experimentally reported values within approximately $0.9$--$2.4\%$ across the four operating points. These bright-port statistics therefore provide an independent cross-check of the physical observable model. By contrast, because $V_{\rm cal}$ is obtained from dark-port decoy statistics, agreement of the dark-port observable cannot itself be regarded as an independent validation metric. Its remaining distance-dependent residual instead identifies the dark-port channel as the principal model-sensitive observable of the present physical description.

Cao \textit{et al.} also reported elevated vacuum-decoy counts in the monitoring basis and discussed reflections within the receiver interferometer as a possible implementation-specific contribution~\cite{cao_experimental_2026}. In their refined analysis, the corresponding monitoring vacuum-decoy contribution was therefore inferred using the experimentally measured $Z$-basis statistics. Such implementation-specific monitoring backgrounds are not explicitly represented in the present optical model and could contribute to the residual sensitivity of the monitoring channel. We therefore do not assign the remaining dark-port discrepancy to a unique physical mechanism.

At 25 km the model overestimates the refined experimentally reported secure-key rate, whereas at 100 km it substantially underestimates it. The large relative residual at 100~km reflects the combined sensitivity of the monitoring-channel model and the nonlinear mapping from the inferred phase-error upper bound $E_p^U$ to the certified secret-key length. We therefore retain this discrepancy as an explicit boundary of the present physical model rather than absorbing it through empirical corrections.

\subsection{Same-Record Comparison of Finite-Key Treatments}
\label{sec:kato_azuma_comparison}

The observable-level architecture enables a controlled comparison of finite-key treatments while holding the physical realization fixed. At each $(L,\mu)$ point, the Kato~\cite{kato_concentration_2020} and Azuma--Hoeffding~\cite{azuma_weighted_1967} evaluators act on the identical detector record $\mathcal O$, so their separation arises solely from the statistical inference.

\begin{figure}[htbp]
    \centering
    \includegraphics[width=0.48\textwidth]{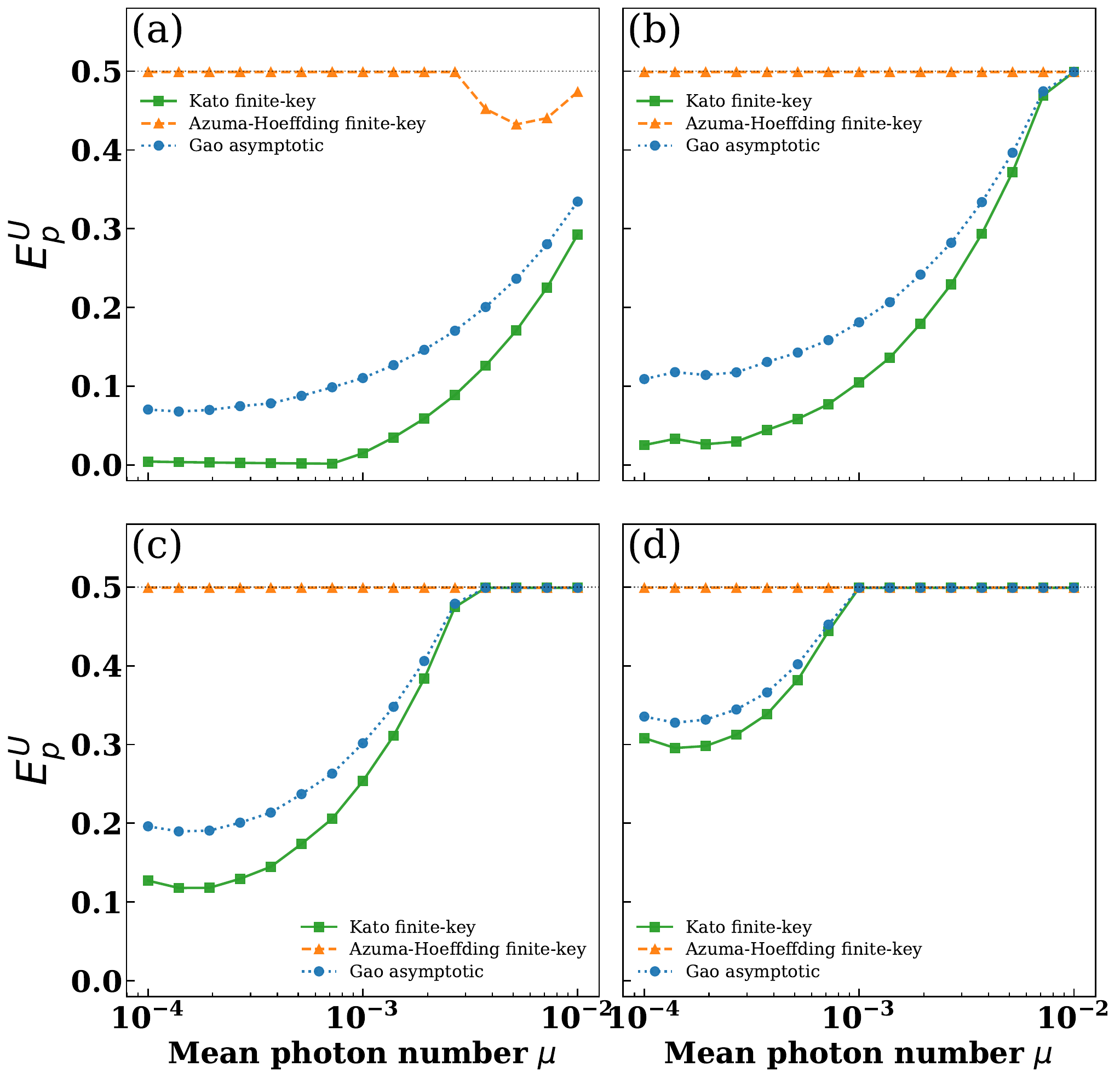}
    \caption{Controlled same-record comparison of finite-key phase-error estimation. The phase-error upper bound $E_p^U$ is evaluated as a function of the coherent-state mean photon number $\mu$ at (a) 25, (b) 50, (c) 75, and (d) 100~km. At each $(L,\mu)$ point, the Kato and Azuma--Hoeffding finite-key evaluators receive the identical detector-resolved record $\mathcal O$; their separation therefore reflects only the statistical security treatment. The Gao asymptotic result is included as a reference.}
\label{fig:kato_azuma_phase_error}
\end{figure}

As shown in Fig.~\ref{fig:kato_azuma_phase_error}, at 25~km the Kato evaluator reaches a minimum phase-error bound
$E_{p,\mathrm{Kato}}^{U}\simeq1.6\times10^{-3}$,
whereas the Azuma--Hoeffding estimate remains above $0.432$. At 50, 75, and 100~km, the corresponding Kato minima are approximately $0.025$, $0.118$, and $0.296$, while the Azuma--Hoeffding bound reaches the $E_p^U=0.5$ security threshold throughout the sampled intensity range for $L\geq50$~km. The increasing Kato bound with distance reflects the progressively sparser monitoring record available for complementary-basis inference. These results therefore demonstrate the practical importance of the count-adaptive treatment for the rare-event records considered here, rather than a universal ordering of the two concentration inequalities. A broader finite-key cross-check is provided in Supplemental Sec.~S1 and Fig.~S2.

\section{Security-Aware Operating-Point Optimization}
\label{sec:optimization}

\subsection{Mean-Photon-Number Optimization}
\label{sec:mu_optimization}

Holding the Kato evaluator fixed, we first optimize the coherent-state mean photon number $\mu$ at a fixed receiver splitting fraction $t_B=0.3$. Interior maxima are obtained at each of the four experimental benchmark distances, with \[\mu^\star= \{5.03\times10^{-3},\, 2.02\times10^{-3},\, 7.88\times10^{-4},\, 3.04\times10^{-4}\}, \] yielding secure-key rates \[ R^\star= \{3.11\times10^{4},\, 4.41\times10^{3},\, 504.5,\, 8.3\}\ {\rm bps} \] at 25, 50, 75, and 100~km, respectively. The complete one-dimensional optimization landscapes are shown in Supplemental Fig.~S2.

The preferred mean photon number decreases systematically with increasing channel loss, falling by more than an order of magnitude between 25 and 100~km. This behavior reflects the competing roles of $\mu$ in COW-QKD: increasing $\mu$ raises the useful detection yield, but also modifies the virtual-state relations entering the complementary-basis inference. The optimum therefore results from a balance between maintaining sufficient detection statistics and retaining a sufficiently tight finite-key phase-error bound.

As shown in Fig.~\ref{fig:optimization_summary}(a), the calibrated model favors somewhat larger intensities than those employed experimentally by Cao \textit{et al.}~\cite{cao_experimental_2026}, with $\mu^\star/\mu_{\rm exp}$ ranging from approximately 1.25 to 1.44 across the four benchmark distances.

\subsection{Joint Source--Receiver Optimization}
\label{sec:joint_optimization}
\begin{figure}[htbp]
    \centering
    \includegraphics[width=0.48\textwidth]{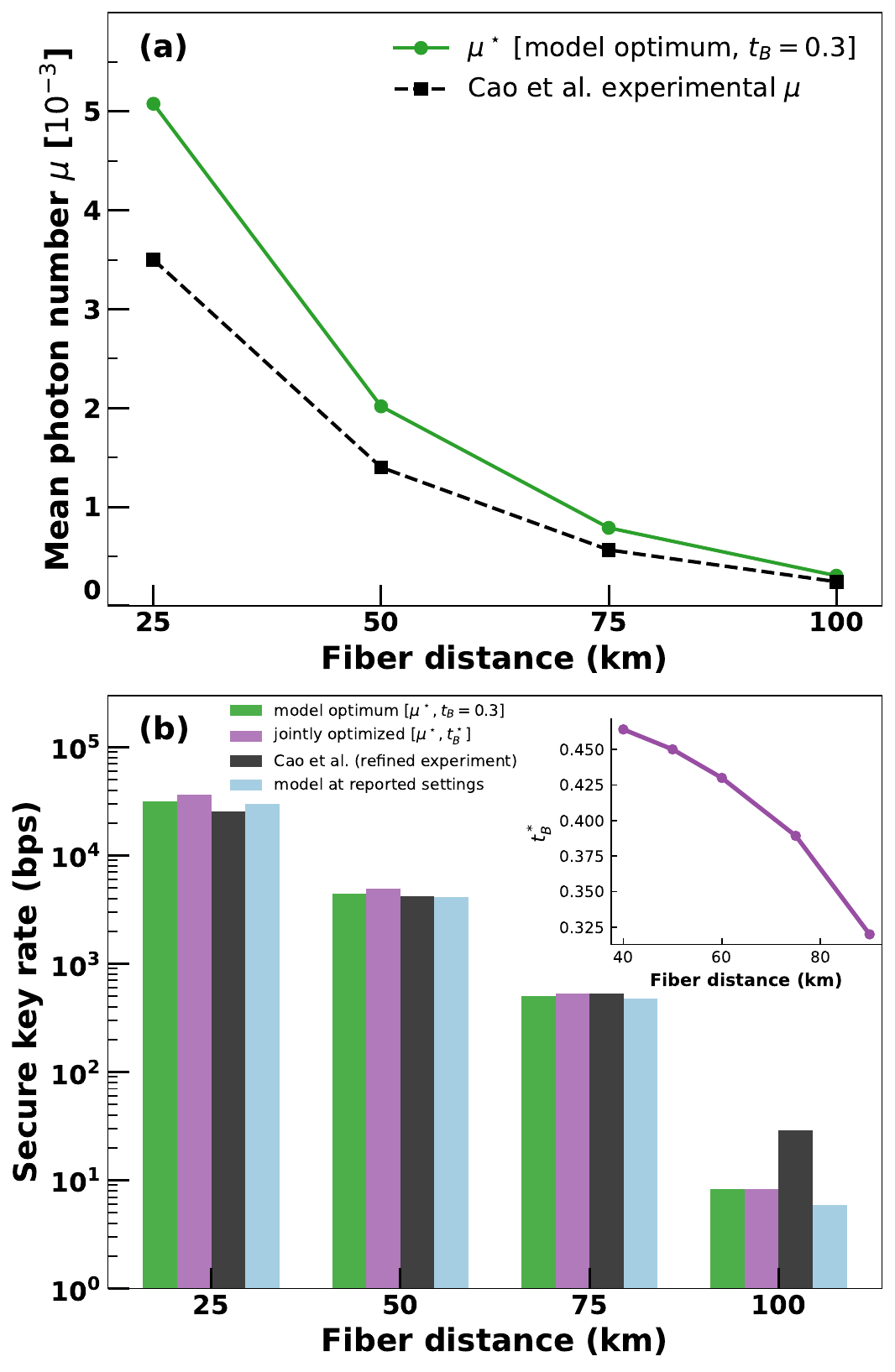}
    \caption{Model-based security-aware operating-point optimization relative to the experimental settings reported by Cao \textit{et al.}~\cite{cao_experimental_2026}. (a) Optimized coherent-state mean photon number $\mu^\star$ obtained by maximizing the calculated Kato finite-key secure-key rate at fixed receiver splitting fraction $t_B=0.3$, compared with the corresponding experimentally employed values of $\mu$ at 25, 50, 75, and 100~km. (b)  Secure-key-rate comparison among the model evaluated at the experimentally reported operating parameters, optimization of $\mu$ alone at fixed $t_B=0.3$, joint optimization of $(\mu,t_B)$, and the refined experimentally reported secure-key rates of Cao \textit{et al.}~\cite{cao_experimental_2026}. Joint optimization increases the model-predicted secure-key rate by approximately $7.4\%$--$20.2\%$ relative to optimization of $\mu$ alone at fixed $t_B=0.3$ across the four benchmark distances. The inset shows the continuously optimized receiver splitting fraction $t_B^\star$ over the 40--90~km range, where it decreases monotonically from $0.464$ to $0.320$ within the present model.}
\label{fig:optimization_summary}
\end{figure}

We next extend the search to joint optimization of $(\mu,t_B)$. Allowing $t_B$ to vary exposes the trade-off between directing received optical power toward data-line detections and retaining sufficient monitoring statistics for coherence estimation. The jointly optimized secure-key rates at the four experimental benchmark distances are shown in Fig.~\ref{fig:optimization_summary}(b). Relative to optimization of $\mu$ alone at fixed $t_B=0.3$, joint optimization increases the model-predicted secure-key rate by \[ \Delta R_{\rm joint}
= \{16.2\%,\,16.5\%,\,7.4\%,\,20.2\%\} \] at 25, 50, 75, and 100~km, respectively.

In addition to these discrete benchmark points, a continuous-distance optimization was performed to examine the evolution of the preferred receiver splitting fraction. Over the 40--90~km range, the optimized value decreases monotonically from $t_B^\star=0.464$ to $t_B^\star=0.320$, as shown in the inset of Fig.~\ref{fig:optimization_summary}(b). This trend indicates that, within the present model, the optimal balance between data acquisition and monitoring changes progressively with channel loss rather than remaining fixed at the experimentally adopted value $t_B=0.3$.

\section{Observable-Level Adversarial Stress Tests}
\label{sec:adversarial_perturbations}

Observable-changing adversarial perturbations are introduced upstream of the security layer, modifying the physical event probabilities and hence the detector record supplied to the otherwise unchanged security evaluator (red dashed pathway in Fig.~\ref{fig:Schematic}). We consider an idealized intercept--resend perturbation~\cite{gisin_towards_2004} and a representative USD-motivated perturbation~\cite{branciard_zero-error_2007} with $q_1=q_3=q$. These examples are used as modular stress tests of the observable-level architecture rather than as complete optimized attack models.

As detailed in the Supplemental Material, the intercept--resend perturbation progressively reduces the monitoring visibility and suppresses the certified key rate (Supplemental Figs.~S3 and S4). The USD-motivated perturbation leaves the data-line QBER comparatively insensitive while modifying the monitoring statistics, thereby increasing the inferred phase-error bound and eventually preventing certification of a positive secret key (Supplemental Figs.~S5 and S6). In both cases, the security degradation arises through the modified detector record rather than through an externally imposed key-rate penalty, illustrating how observable-changing adversarial mechanisms can be incorporated within the same simulation--security interface.

\section{Conclusions}
\label{sec:conclusion}

We have developed an observable-level modular framework for
finite-key analysis of Coherent-One-Way (COW) quantum key
distribution, organized around the interface
$\boldsymbol{\Theta}_{\rm phys}
\xrightarrow{\mathcal P}\mathbf p
\xrightarrow{\mathcal G}\mathcal O
\xrightarrow{\mathcal S_j}\mathcal S_j(\mathcal O)$.
By separating memoryless physical event probabilities, finite-block detector records, and statistical security inference, the framework allows the photonic model, count-generation layer, and security evaluator to be examined and modified independently.

The framework was benchmarked against the experimental operating points of Cao \textit{et al.}~\cite{cao_experimental_2026}. Simulated monitoring bright-port decoy counts agree with the reported values within approximately 0.9\%--2.4\%, while the Kato finite-key rates differ from the refined experimentally reported secure-key rates by approximately $-1.5\%$ at 50~km and $-10.3\%$ at 75~km. The remaining distance-dependent dark-port residual identifies the principal model-sensitive observable of the present physical description and is retained explicitly rather than absorbed through empirical correction.

Applying Kato and Azuma--Hoeffding concentration treatments to identical detector records isolates the effect of finite-key statistical inference from the underlying physical simulation. Holding the Kato evaluator fixed, joint optimization of the coherent-state mean photon number $\mu$ and receiver splitting fraction $t_B$ increases the model-predicted secure-key rate by $7.4\%$--$20.2\%$ relative to optimizing $\mu$ alone at fixed $t_B=0.3$.

Representative intercept--resend and USD-motivated perturbations further demonstrate that observable-changing mechanisms can be introduced upstream of the security layer and propagated through the same detector-record interface without modifying the downstream evaluator. The principal contribution of this work is therefore not a new COW security proof, but a structured computational interface connecting physical modeling, finite-block experimental statistics, and interchangeable security inference.

The numerical implementation combines continuous-variable Gaussian propagation, deterministic Gauss--Hermite phase averaging, an effective non-paralyzable detector-recovery model, aggregate binomial count generation, virtual-state Cauchy--Schwarz inference, and interchangeable finite-key concentration treatments within the same observable interface. Because these elements are separated by explicit interfaces, individual physical or statistical components can be refined or replaced without redesigning the complete computational pipeline. This architecture provides a basis for experimentally grounded protocol comparison, security-aware parameter optimization, and future extensions incorporating more detailed detector dynamics and alternative security models.

\section*{Acknowledgments}

The authors gratefully acknowledge the valuable help and support of Manomay Tyagi, Ridham Mahajan, and Prof. C. S. Yadav. A.D.L. and U.G. acknowledge financial support from the Department of Biotechnology (DBT), GOI, under Project Grant No.~BT/PR65032/QTBT/176/26/2025 and from the Indian Institute of Technology Mandi (IIT Mandi) Seed Grant Project No.~IITM/SG/ADL/173. M.C. acknowledges financial support from the IIT Mandi HTRA Fellowship.

\section*{Data Availability}

The numerical data and simulation code supporting the findings of this study are openly available in the Zenodo repository at
\href{https://doi.org/10.5281/zenodo.23456644}{10.5281/zenodo.23492296}.
The actively maintained development version of the software is available in the
\href{https://github.com/Aryan-K1509/Observable-Level-Framework-for-Finite-Key-Security-in-COW-QKD}{GitHub repository}.

\bibliographystyle{unsrt}
\bibliography{Refs}

\end{document}


\title{Supplemental Material for: Observable-Level Framework for Finite-Key Security in Coherent-One-Way Quantum Key Distribution}

\author{Aryan Kuntal}
\email{aryan.kuntal09@gmail.com}
\affiliation{Center for Quantum Science and Technologies, Indian Institute of Technology Mandi, Kamand 175075, Himachal Pradesh, India}
\affiliation{Fergusson College, Gopal Krishna Gokhale Road, Shivajinagar, Pune 411004, Maharashtra, India}

\author{Utkarsh Giradkar}
\email{utkarshgiradkar1729@gmail.com}
\affiliation{Center for Quantum Science and Technologies, Indian Institute of Technology Mandi, Kamand 175075, Himachal Pradesh, India}

\author{Mudit Chaudhary}
\affiliation{Center for Quantum Science and Technologies, Indian Institute of Technology Mandi, Kamand 175075, Himachal Pradesh, India}
\affiliation{School of Physical Sciences, Indian Institute of Technology Mandi, Kamand 175075, Himachal Pradesh, India}

\author{Amit D. Lad}
\email{amitdlad@iitmandi.ac.in}
\affiliation{Center for Quantum Science and Technologies, Indian Institute of Technology Mandi, Kamand 175075, Himachal Pradesh, India}

\maketitle

\setcounter{figure}{0}
\setcounter{equation}{0}
\renewcommand{\thefigure}{S\arabic{figure}}
\renewcommand{\theequation}{S\arabic{equation}}

This Supplemental Material provides implementation cross-checks and extended numerical results supporting the observable-level framework presented in the main text. Section~S1 benchmarks the finite-key implementation under the parameter and block-size conventions of Li \textit{et al.}~\cite{li_finite-key_2024}; Sec.~S2 presents the mean-photon-number optimization landscapes; Sec.~S3 examines the representative intercept--resend perturbation together with an observable-level monitoring-visibility verification; and Sec.~S4 analyzes the representative USD-motivated perturbation through its effects on the certified secure-key rate and the finite-key phase-error bound. Unless otherwise stated, the notation, detector conventions, and physical parameters follow those defined in the main text.

\subsection*{S1. Finite-key cross-check under benchmark conditions}

\begin{figure}[htbp]
    \centering
    \includegraphics[width=0.6\columnwidth]{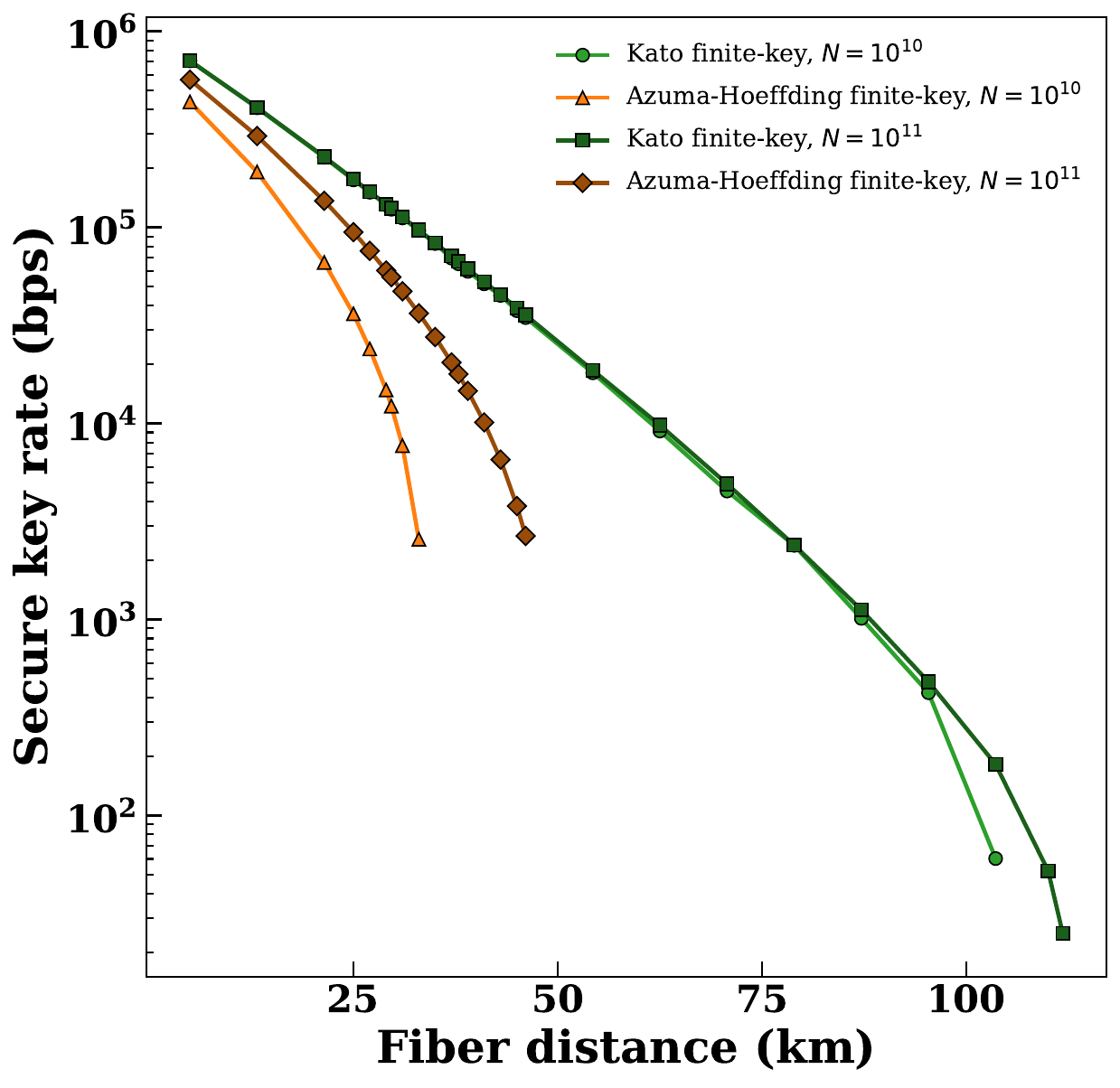}
    \caption{Finite-key secure-key-rate comparison between the Kato and Azuma--Hoeffding concentration treatments under benchmark conditions following Li \textit{et al.}~\cite{li_finite-key_2024}. Results are shown for block sizes $N=10^{10}$ and $10^{11}$ using the block-size convention of Ref.~\cite{li_finite-key_2024}. The comparison provides a qualitative implementation cross-check and illustrates the reduced statistical penalty obtained with the count-adaptive Kato treatment for the low-count monitoring regime considered here.}
    \label{fig:supp_li_crosscheck}
\end{figure}

As an implementation-level cross-check, we reproduce a representative parameter regime and block-size convention from Li \textit{et al.}~\cite{li_finite-key_2024} and evaluate the Kato and Azuma--Hoeffding treatments within our implementation. Figure~\ref{fig:supp_li_crosscheck} shows the resulting secure-key rate as a function of fiber distance for block sizes $N=10^{10}$ and $10^{11}$.

The two concentration treatments exhibit markedly different finite-size
behavior as the monitoring statistics become sparse. The Azuma--Hoeffding evaluator produces a larger statistical penalty, whereas the count-adaptive Kato treatment retains tighter bounds. Increasing the block size reduces the finite-size penalty for both methods and extends the distance over which a positive secret key can be certified. This comparison is intended as a qualitative implementation cross-check rather than as a universal ordering of the two concentration inequalities, and is consistent with the behavior reported by Li \textit{et al.}~\cite{li_finite-key_2024}.

\subsection*{S2. Mean-photon-number optimization landscape}

\begin{figure}[htbp]
    \centering
    \includegraphics[width=0.6\columnwidth]{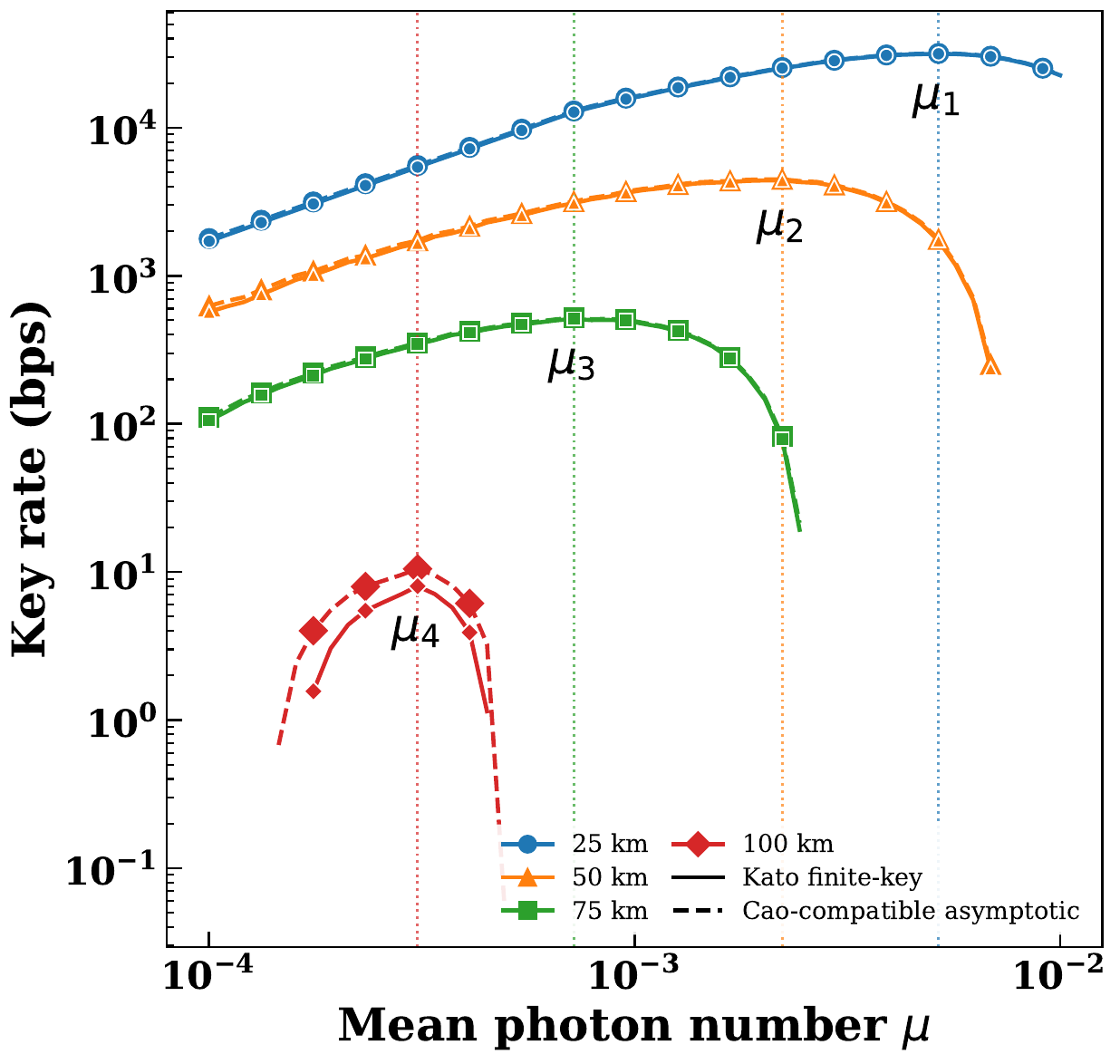}
    \caption{Kato finite-key secure-key rate as a function of the coherent-state mean photon number $\mu$ at 25, 50, 75, and 100~km, with the receiver splitting fraction fixed at $t_B=0.3$. Each curve exhibits an interior maximum, with the preferred intensity shifting toward smaller $\mu$ as the channel loss increases. The vertical dotted lines mark the optimized values $\mu^\star$ for the corresponding transmission distances. The Cao-compatible asymptotic calculation is included as a reference.}
    \label{fig:supp_mu_landscape}
\end{figure}

To resolve the intensity dependence underlying the operating-point optimization in the main text, we evaluate the Kato finite-key secure-key rate as a function of the coherent-state mean photon number $\mu$ with the receiver splitting fraction fixed at $t_B=0.3$. Figure~\ref{fig:supp_mu_landscape} shows the resulting landscapes at 25, 50, 75, and 100~km using the detector parameters corresponding to the Cao \textit{et al.} operating points~\cite{cao_experimental_2026}.

The corresponding optima are $\mu^\star=\{5.03\times10^{-3},\,2.02\times10^{-3},\, 7.88\times10^{-4},\,3.04\times10^{-4}\}$, giving secure-key rates of approximately $\{3.11\times10^{4},\,4.41\times10^{3},\,504.5,\,8.3\}$~bps at 25, 50, 75, and 100~km, respectively. The decrease of $\mu^\star$ with distance reflects the balance between maintaining sufficient detection yield and preserving sufficiently favorable monitoring statistics for the finite-key phase-error estimate. These one-dimensional optima provide the fixed-$t_B$ reference used for the joint $(\mu,t_B)$ optimization discussed in the main text.

\FloatBarrier
\subsection*{S3. Intercept--resend perturbation of monitoring coherence}

To test the propagation of an observable-changing perturbation through the simulation pipeline, we introduce an idealized intercept--resend mechanism parameterized by the intercepted fraction $p_{\rm IR}$. The perturbation is applied upstream of the security evaluator so that its effect enters exclusively through the modified detector statistics. This construction is used as a representative stress test of the observable-level architecture rather than as a complete optimized intercept--resend attack model.

\begin{figure}[htbp]
    \centering
    \includegraphics[width=0.6\columnwidth]{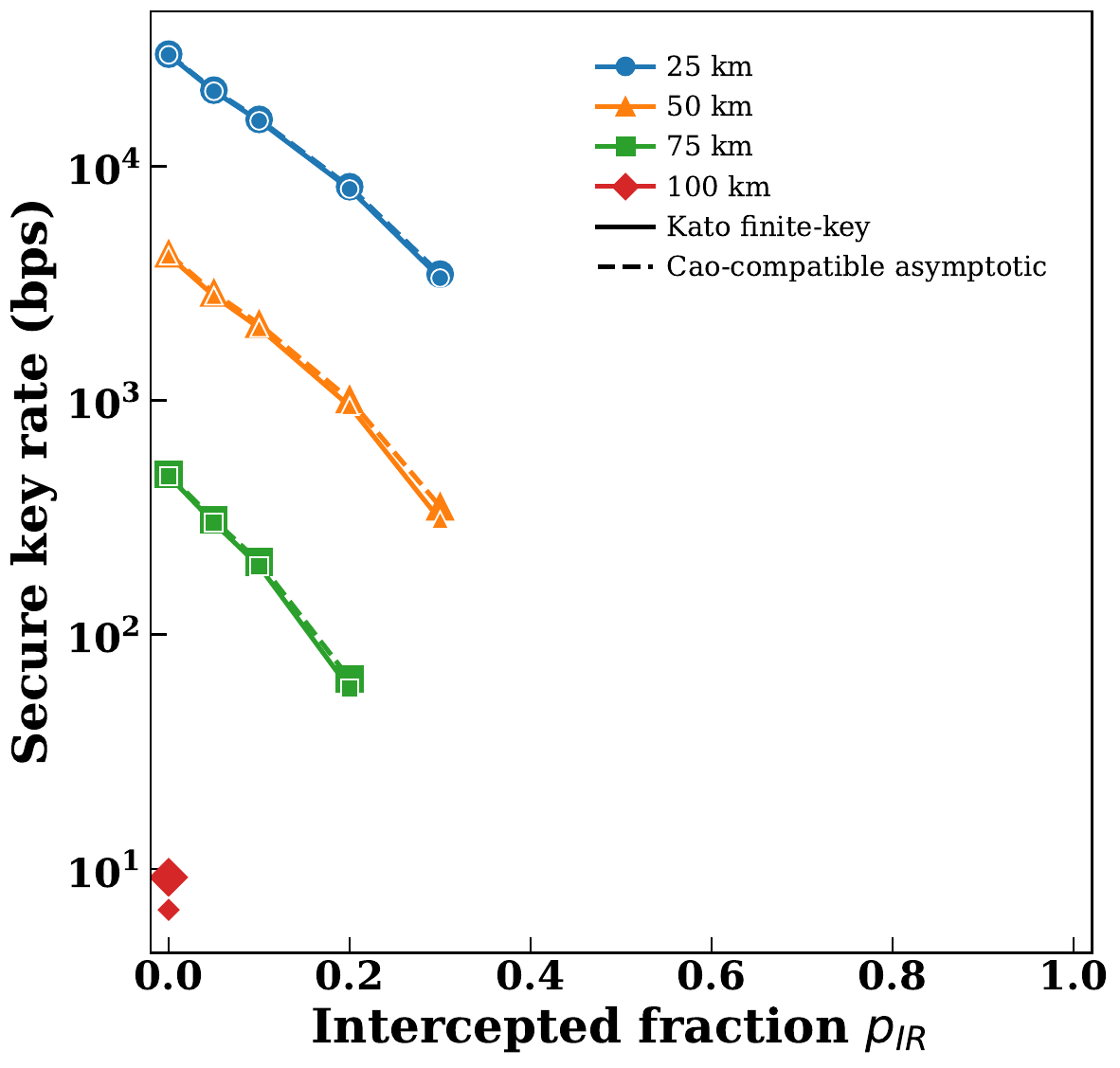}
    \caption{Secure-key rate under the representative intercept--resend perturbation as a function of the intercepted fraction $p_{\rm IR}$. Results are shown for transmission distances of 25, 50, 75, and 100~km. The solid curves denote the Kato finite-key evaluation, while the corresponding dashed curves show the Cao-compatible asymptotic reference. Increasing $p_{\rm IR}$ progressively modifies the monitoring statistics and suppresses the certified secure-key rate. Only positive certified rates are displayed on the logarithmic vertical scale; each curve terminates when the corresponding security treatment no longer yields a positive key rate.}
    \label{fig:IR_keyrate}
\end{figure}

Fig.~\ref{fig:IR_keyrate} shows the resulting secure-key rate as $p_{\rm IR}$ is increased from zero to unity. The degradation of inter-pulse coherence modifies the monitoring statistics and therefore increases the finite-key security penalty.
The effect becomes progressively more restrictive with increasing channel loss: the shorter-distance configurations retain a positive certified key over a wider range of $p_{\rm IR}$, whereas the higher-loss configurations become non-key-generating at smaller perturbation strengths.

\begin{figure}[htbp]
    \centering
    \includegraphics[width=0.6\columnwidth]{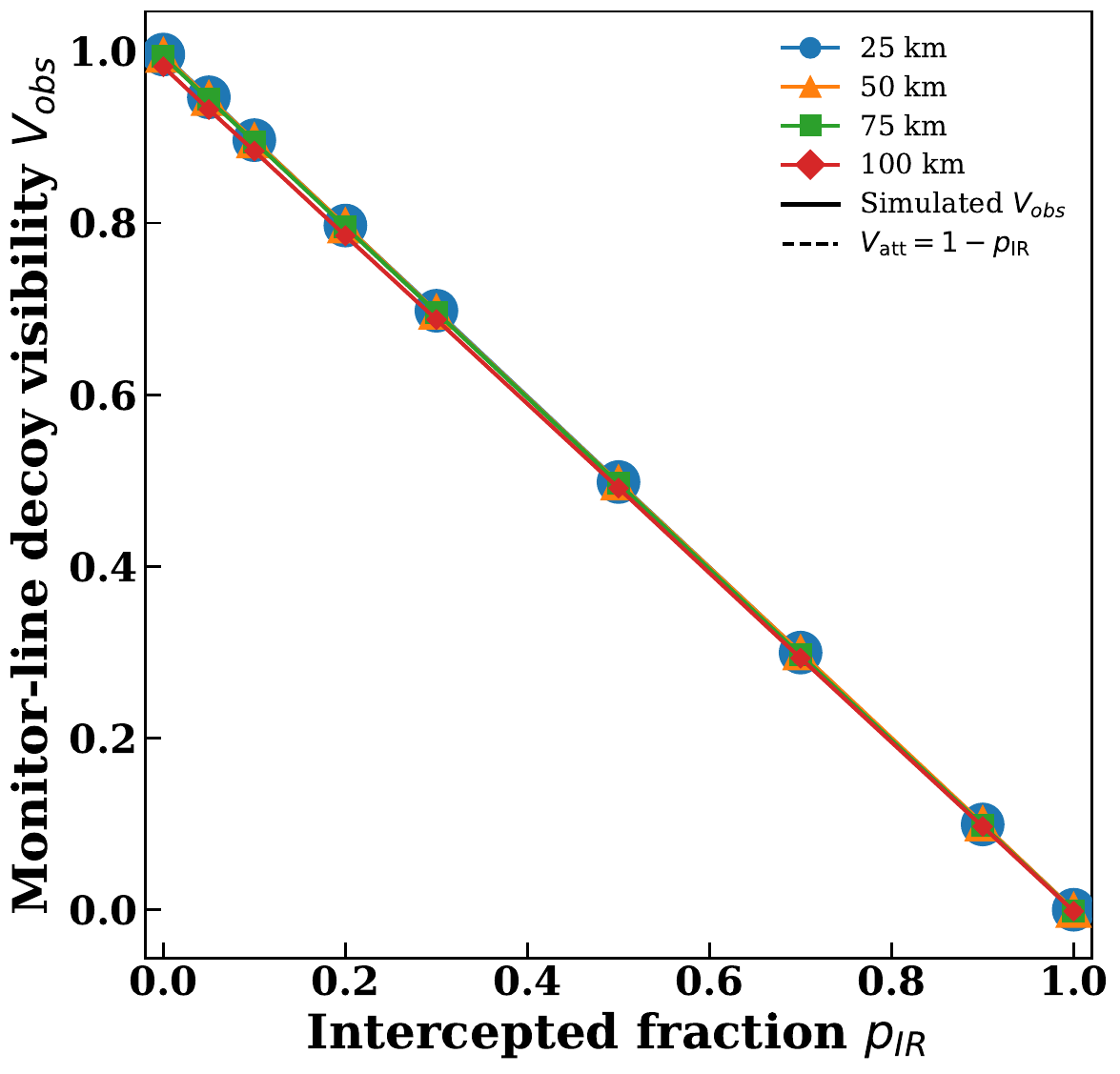}
    \caption{Observable-level verification of the representative intercept--resend perturbation. The monitoring-line decoy visibility reconstructed from the simulated detector counts is compared with the imposed relation $V_{\rm att}=1-p_{\rm IR}$ for 25, 50, 75, and 100~km. Agreement within numerical precision verifies the implementation of the perturbation adapter in the physical-to-observable pipeline.}
    \label{fig:supp_ir_visibility}
\end{figure}

For the adopted perturbation model, the imposed monitoring visibility is
\begin{equation}
V_{\rm att}=1-p_{\rm IR}.
\end{equation}
Figure~\ref{fig:supp_ir_visibility} compares this imposed relation with the visibility reconstructed from the simulated monitoring-detector counts. The reconstructed values follow the expected dependence within numerical precision across all four transmission distances, verifying that the perturbation introduced at the physical layer is transferred consistently to the observable detector record.

\FloatBarrier
\subsection*{S4. Representative USD-motivated perturbation}

We next consider a representative perturbation motivated by unambiguous-state-discrimination (USD) attacks on COW-QKD \cite{branciard_zero-error_2007}. Following the parametrization used in the main text, the perturbation strengths are taken as $q_1=q_3=q$, with $q$ varied from zero to unity. This construction is used as an observable-level stress test rather than as a complete optimization over all physically allowed USD strategies.

\begin{figure}[htbp]
    \centering
    \includegraphics[width=0.6\columnwidth]{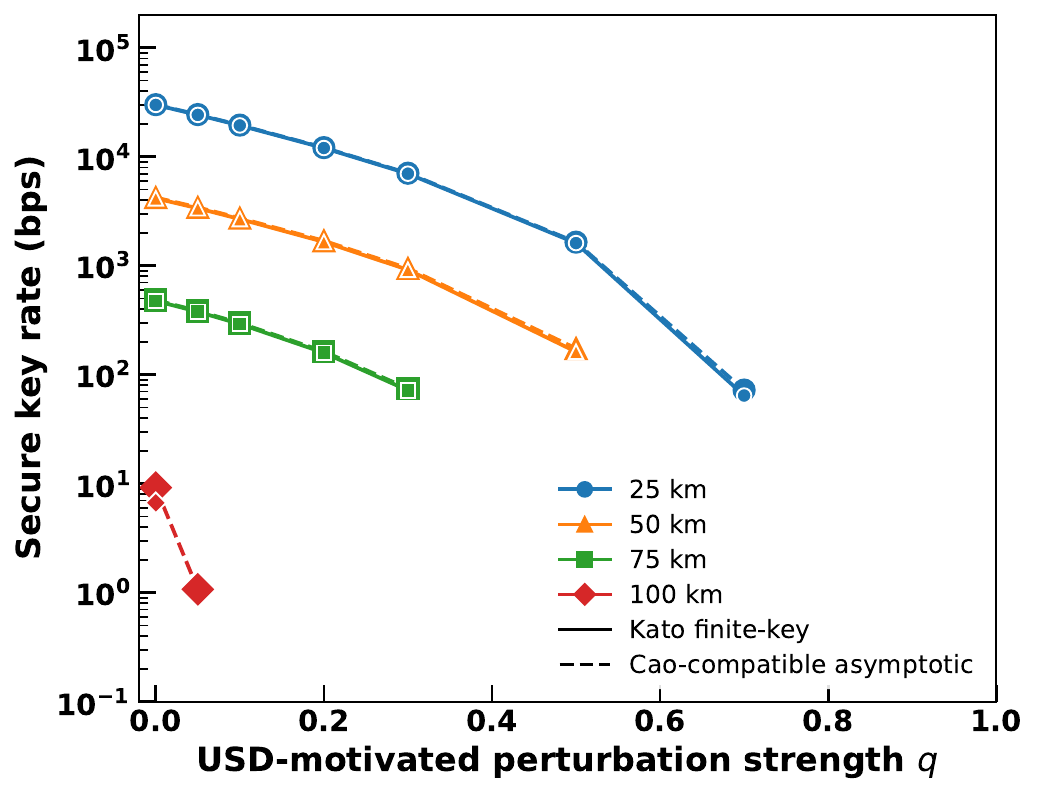}
    \caption{Secure-key rate under the representative USD-motivated perturbation as a function of the perturbation strength $q$, with $q_1=q_3=q$. Results are shown at transmission distances of 25, 50, 75, and 100~km. The solid curves denote the Kato finite-key evaluation, while the corresponding dashed curves show the Cao-compatible asymptotic reference. At each perturbation strength, the modified detector record is propagated through the otherwise unchanged security evaluator. Increasing $q$ modifies the monitoring statistics and progressively suppresses the certified secure-key rate, with higher-loss links becoming non-key-generating at smaller values of $q$. Only positive rates are displayed on the logarithmic vertical scale.}
    \label{fig:supp_usd_keyrate}
\end{figure}

\begin{figure} [htbp]
    \centering
    \includegraphics[width=0.6\columnwidth]{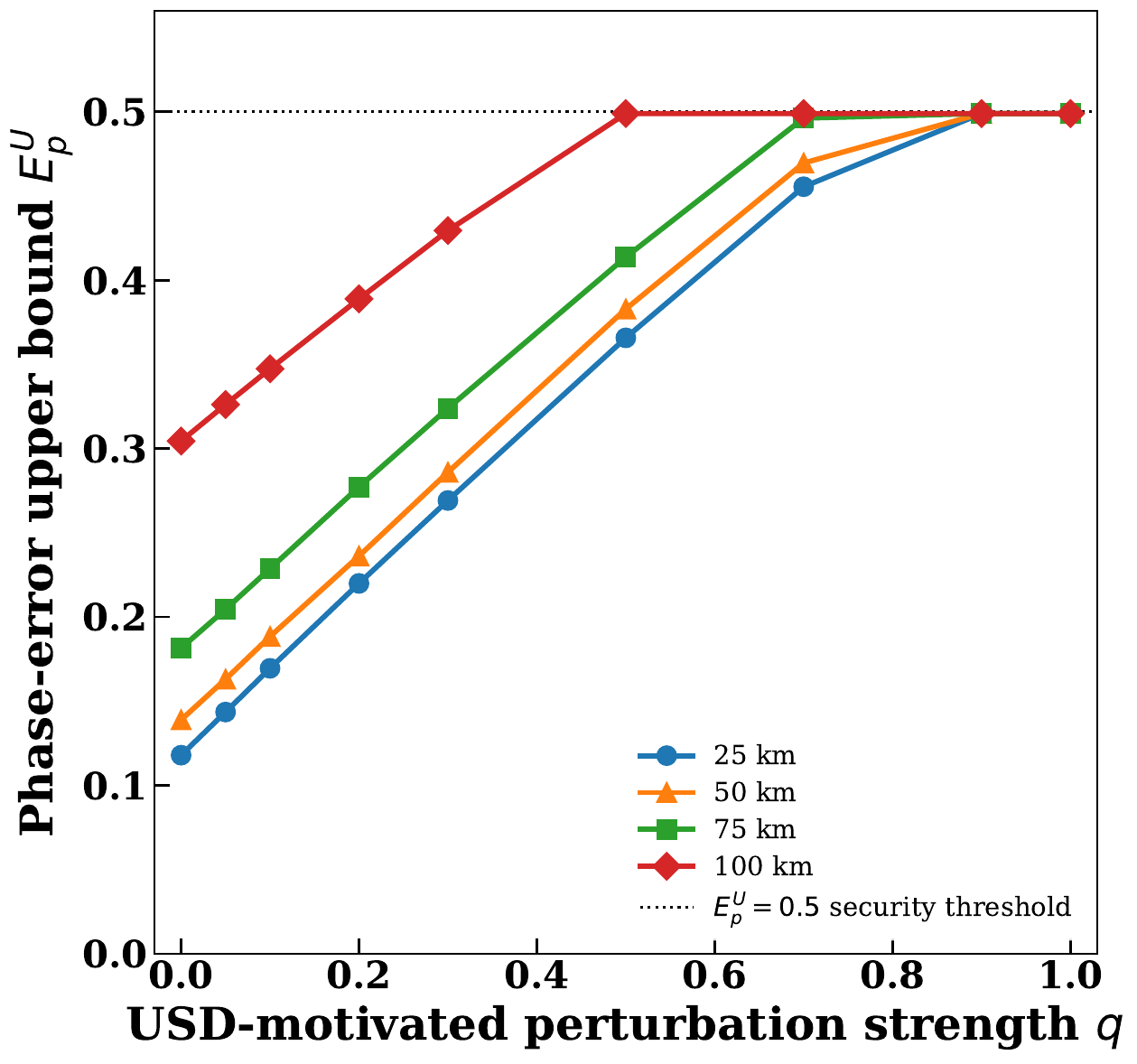}
    \caption{Finite-key phase-error upper bound $E_p^U$ as a function of the representative USD-motivated perturbation strength $q$ at 25, 50, 75, and 100~km. Increasing $q$ modifies the monitoring statistics and drives $E_p^U$ toward the security threshold $E_p^U=1/2$. The horizontal dashed line marks this threshold. Once it is reached, the privacy contribution $1-h_2(E_p^U)$ vanishes and no positive secret key can be certified by the adopted finite-key security bound.}
    \label{fig:supp_usd_phase}
\end{figure}

The perturbation is designed to leave the data-line statistics comparatively insensitive while modifying the inter-pulse coherence probed by the monitoring line. Consequently, increasing $q$ primarily changes the detector statistics entering the complementary-basis inference and therefore increases the inferred phase-error upper bound.

Figure~\ref{fig:supp_usd_keyrate} shows the corresponding secure-key rates. As $q$ increases, the certified rate decreases and eventually vanishes, with the cutoff occurring at progressively smaller perturbation strength as the channel loss increases. For the parameter sets considered here, the 100~km configuration becomes non-key-generating for $q\gtrsim0.1$, whereas the shorter-distance configurations tolerate progressively larger perturbation strengths before the finite-key rate reaches zero.

The origin of this behavior is shown directly in Fig.~\ref{fig:supp_usd_phase}. The perturbation degrades the monitoring statistics entering the virtual-state inference, causing the finite-key phase-error upper bound $E_p^U$ to increase with $q$. Once $E_p^U$ reaches $1/2$, the privacy term $1-h_2(E_p^U)$ in Eq.~(5) of the main text vanishes, and the adopted finite-key expression can no longer certify a positive secret key.

The loss of a positive certified key at $E_p^U=1/2$ should not be interpreted as implying that the eavesdropper necessarily possesses complete knowledge of the raw key. Rather, the available detector record and the adopted finite-key security bound are no longer sufficient to establish a positive secret-key length. Together, Figs.~\ref{fig:supp_usd_keyrate} and \ref{fig:supp_usd_phase} demonstrate that an adversarial perturbation acting upstream of the security layer can be propagated through the same observable interface without introducing an external attack-dependent penalty into the key-rate expression.
\FloatBarrier

\bibliographystyle{unsrt}
\bibliography{Refs}